\documentclass[nofootinbib,superscriptaddress,showpacs,twocolumn]{revtex4}
\usepackage{amssymb}
\usepackage{amsmath}
\usepackage{txfonts}
\usepackage{mathdots}
\usepackage[colorlinks]{hyperref} 
\usepackage{breqn} 
\usepackage[classicReIm]{kpfonts}
\usepackage{comment}   
\usepackage{multirow} 
\usepackage{longtable}
\usepackage{lipsum} 
\usepackage{adjustbox}
\usepackage{tabularx}
\usepackage{lipsum}
\usepackage{titlesec}
\usepackage{graphicx}
\usepackage{capt-of}
\usepackage{float}
\renewcommand{\figurename}{Fig.}

\begin{document}
\title{{\bf  Thomas Fermi Screening Length in $q$-Deformed Statistical Mechanics}}
\author{\bf M. Mohammadi Sabet\footnote{Corresponding author email: m.mohamadisabet@ilam.ac.ir}}
\affiliation{Physics Department; Faculty of science; Ilam university; Ilam; Iran}
\begin{abstract}
The $q$-deformed statistical mechanics for fermions has been used to investigate the Thomas-Fermi screening length at finite temperature. Considering linear response, the calculations have been made 
at weakly nondegenerate regime. The results show that $q$-deformation has significance effects on screening length at higher temperatures. It is also shown that the $q$-deformation effects vanish at zero temperature limit. One can find that more correction terms of deformation have more effects on screening length. The bahaviour of screening length is different for different values of $q$.
\end{abstract}
\maketitle
{\bf Keywords: } 
\section{Introduction}
\label{Intro}

Thomas-Fermi model is a semi-classical one provided to investigate the many body effects in quantum systems.
This model is applicable in  metals, solid state physics, atomic physics, as well as astrophysics \cite{spruch, slater}.
In spite of its crudity, because of creating qualified views, this model is used in many fields of modern physics some of which are regarded its generalization \cite{shababi,kamel}. In this model, electrons are considered as a homogeneous uniform electron gas with charge density $-n_0 e$, obeying Fermi-Dirac statistics, superimposed on a background of positive charge density $n_0 e$. 

Considering a point charge $Q$ in a sea of such electrons, Thomas-Fermi model leads to screening in the coulomb potential $\phi$ expressed by the Thomas-Fermi equation\cite{Shivamoggy}
\begin{equation}\label{TF_model_1}
\nabla^2 \phi(\mathbf{r}) = 4 \pi e \left( n(\mathbf{r}) - n_0 \right)-4 \pi Q \delta(\bf{r})	
\end{equation}
The Thomas-Fermi screening length, \( \lambda_{F} \), is a characteristic length scale that describes how quickly the electron density as well as the associated electrostatic potential decay away from a charged object or within a material. It quantifies the range over which the electron density screens the electrostatic potential due to the presence of other charged particles. 

On the other hand, it is found that there is some classes of systems that ordinary quantum statistical mechanics, known as Gibss-Blotzman statistical mechanics, may not be proper \cite{Borges,Sotolongo,Soonmin} and  a kind of extension is needed to describe these systems. Regarding this issue, two principal methods have been proposed for intermediate statistics: the nonextensive statistics \cite{Tsallis} and q-deformed theory \cite{Biedenharn,Macfarlane}. Because of its possibility to apply in different field of physics such as anyon physics \cite{Caracciolo,Lerda}, thermodynamics of ideal Fermi gas \cite{Hui,Cai} and references therein, regarding other fields of physics, the q-deformed statistics has taken great interest last decade. In this regard, it is shown that application of q-deformation in fermion systems changes its thermodynamics properties.

Upton the foregoing discussion,  the effect of q-deformed Fermi-Dirac statistics on the Thomas-Fermi screening length has been investigated in this paper. The paper is organized as follows. In the next section q-deformed algebra and statistical mechanics are briefly introduced. Then the this intermediate statistics will be used in the Thomas-Fermi equation to investigate the effect of q-deformation on the screening length. Finally, results and discussion are presented.

\section{q-deformed algebra and statistical distribution function for fermions}
The symmetric $q$-deformed fermion algebra is defined as follows,
\begin{equation}
\left[\hat{N}, \hat{a}^{\dagger}\right]=\hat{a}^{\dagger}, \quad[\hat{N}, \hat{a}]=-\hat{a},
\end{equation}
and
\begin{equation}
\hat{a}^{\dagger} \hat{a}=[\hat{N}], \quad \hat{a} \,\hat{a}^{\dagger}=[1-\hat{N}],
\end{equation}
where $\hat{a}^{\dagger}$, $\hat{a}$ and $\hat{N}$ are creation, annihilation and number operator, respectively and the $q$-basic number $[x]$ is defined as
\begin{equation}
[x]=\frac{q^x-q^{-x}}{q-q^{-1}}	
\end{equation}
Here $q$ is the deformation parameter. The Hilbert space of $q$-deformed fermions with the basis $|n\rangle$ is as \cite{}
\begin{equation}
\begin{aligned}
	& \hat{N}|n\rangle=n|n\rangle, \quad \hat{a}|0\rangle=0, \\
	& \hat{a}^{+}|n\rangle=[1-n]^{1 / 2}|n+1\rangle, \\
	& \hat{a}|n\rangle=[n]^{1 / 2}|n-1\rangle .
\end{aligned}
\end{equation}
The eigenvalues of number operator $\hat{N}$ take only the values $0$ and $1$ and the Pauli principle is satisfied in $q$-deformed fermions.
The mean value of the $q$-deformed occupation number is defined by \cite{Cai}
\begin{equation}
	\left[f_{k, q}\right]=\frac{1}{\Xi} \operatorname{tr}\left\{\exp (-\beta \hat{H})\left[\hat{N}_k\right]\right\},
\end{equation}
where $\beta=1/k_BT$, $\hat{H}$ is the Hamiltonian
\begin{equation}
\hat{H}=\sum_k\left(\varepsilon_k-\mu\right) \hat{N}_k,
\end{equation}
and  $k$ is a state label, $\hat{N}_k$ and $\varepsilon_k$ are the number operator and energy associated with state $k$, respectively, and $\mu$ is the chemical potential. Following \cite{Cai} the statistical distribution function of the $q$-deformed fermions can be derived as
\begin{equation}\label{distfunction1}
	f_{k, q}=\frac{1}{2 \ln q} \ln \left[\frac{z^{-1} \exp \left(\beta \varepsilon_k\right)+q}{z^{-1} \exp \left(\beta \varepsilon_k\right)+q^{-1}}\right]
\end{equation}
where $z=exp(\beta\mu)$ is the fugacity of the system. One important properties of this distribution function is $f_{k,q}=f_{k,1/q}$. One can simply prove that when $q=1$ we have the standard Fermi-Dirac distribution and the $q-$deformed fermions are the same as ordinary fermions,
\begin{equation}
	f_{k,q}=\frac{1}{z^{-1}exp(\beta \varepsilon_k)+1}
\end{equation} 
According to Eq.(\ref{distfunction1}), total number of particles, $N$ and total energy of system $U$ can be, receptively, given by
\begin{equation}\label{totalnumber1}
	N=\sum_k\frac{1}{2 \ln q} \ln \left[\frac{z^{-1} \exp \left(\beta \varepsilon_k\right)+q}{z^{-1} \exp \left(\beta \varepsilon_k\right)+q^{-1}}\right]
\end{equation}
and
\begin{equation}\label{totalenergy1}
U=\sum_k\frac{\epsilon_k}{2 \ln q} \ln \left[\frac{z^{-1} \exp \left(\beta \varepsilon_k\right)+q}{z^{-1} \exp \left(\beta \varepsilon_k\right)+q^{-1}}\right]	
\end{equation}
In the large particle number limit, the sum over $k$ replaced by integration and therefore Eqs \eqref{totalnumber1} and \eqref{totalenergy1} can be rewritten as,
\begin{equation}\label{totalnumber2}
	 \begin{aligned}
	N& =\frac{g}{h^3}\int d\vec{p}d\vec{x}\frac{1}{2Lnq}\ln \left[\frac{z^{-1} \exp \left(\beta \varepsilon(p)\right)+q}{z^{-1} \exp \left(\beta \varepsilon(p)\right)+q^{-1}}\right] \\
	& =\frac{gV}{\lambda^3}h_{3/2}(z,q) 
\end{aligned}
\end{equation}
and
\begin{equation}\label{totalenergy2}
	 \begin{aligned}
	U &=\frac{g}{h^3}\int d\vec{p}d\vec{x}\frac{\varepsilon(p)}{2Lnq}\ln \left[\frac{z^{-1} \exp \left(\beta \varepsilon(p)\right)+q}{z^{-1} \exp \left(\beta \varepsilon(p)\right)+q^{-1}}\right] \\
	&=\frac{3}{2}k_BT\frac{gV}{\lambda^3}h_{5/2}(z,q) 
	 \end{aligned}
\end{equation}
where $V$ is the volume, $\varepsilon(P)=p^2/2m$ and $\lambda$ is thermal wavelength,
\begin{equation}
	\lambda=\frac{h}{\sqrt{2 \pi m k_{\mathrm{B}} T}}
\end{equation}
In Eqs \eqref{totalnumber2} and \eqref{totalenergy2}, $h_n(z,q)$ is the generalized Fermi integral of q-fermions
\begin{equation}\label{hn}
h_n(z, q)=\frac{1}{\Gamma(n)} \int_0^{\infty} \mathrm{d} x x^{n-1} \frac{1}{2 \ln q} \ln \left[\frac{z^{-1} \exp (x)+q}{z^{-1} \exp (x)+q^{-1}}\right]	
\end{equation}
and $\Gamma(x)=$ $\int_0^{\infty} \exp (-t) t^{x-1} \mathrm{~d} t$ is the Gamma function. It can be found that when $q=1$, Eq. \eqref{hn} is just the standard Fermi integral,
\begin{equation}
	h_n(z,q)=\frac{1}{\Gamma(n)} \int_{0}^{\infty}dx\frac{x^{n-1}}{z^{-1}exp(x)+1}
\end{equation}
\section{Thomas-Fermi screening length in the q-deformed statistical mechanics}\label{sec:gen}

In order to investigate the $q$-deformation effects on the Thomas-Fermi screening length, the fugacity, $z=exp(\beta \mu)$ is considered as $\tilde{z}=exp(\beta (\mu-e\phi))$. Therefore, the number density, $n$, in Eq. \eqref{TF_model_1} is as follows,
\begin{equation}
	 \begin{aligned}
		n& =\frac{g}{h^3}\int d\vec{p}\frac{1}{2Lnq}\ln \left[\frac{exp(-\beta (\mu-e\phi)) \exp \left(\beta \varepsilon(p)\right)+q}{exp(-\beta (\mu-e\phi)) \exp \left(\beta \varepsilon(p)\right)+q^{-1}}\right] \\
		& =\frac{g}{\lambda^3}h_{3/2}(\tilde{z},q),
	\end{aligned}
\end{equation}
Hence, Eq. \eqref{TF_model_1} is rewritten as 
\begin{equation}\label{TF_model_2}
	\begin{aligned}
		\nabla^2 \phi(\mathbf{r})& = 4 \pi e \frac{g}{h^3} \\
		&=\left\{%
		\int d\vec{p}\frac{1}{2Lnq}%
		\left(%
		 \ln \left[\frac{exp(\beta (\mu-e\phi)) \exp \left(\beta \varepsilon(p)\right)+q}{exp(-\beta (\mu-e\phi)) \exp \left(\beta \varepsilon(p)\right)+q^{-1}}\right]
		 \right. \right.\\
		&	%
		\left. \left.	-\ln \left[\frac{exp(-\beta (\mu-\varepsilon(p)))+q}{exp(-\beta (\mu-\varepsilon(p)) +q^{-1})}\right]%
			\right) \right\} \\ %
		&-4 \pi Q \delta(\bf{r}). \\
	\end{aligned}	
\end{equation}
Now using Eq. \eqref{totalnumber2} the Thomas-Fermi model takes the following form,
\begin{equation}\label{TF3}
	\nabla^2 \phi(\mathbf{r})=4 \pi e \frac{g}{h^3 \lambda^3}\left\{h_{3/2}(\tilde{z},q)-h_{3/2}(z,q)\right\}-4 \pi Q \delta(\bf{r}) 
\end{equation}
In the weakly nondegenerate case, i.e., $|\beta \tilde{\mu}|=|\beta (\mu-e\phi)|\gg1$ as well as $|\beta \mu|\gg1$, the generalized Fermi integral (of $q$-fermions) can be written as
\begin{equation}
	\begin{aligned}
		h_n(z, q)=\frac{(\ln z)^n}{\Gamma(n)}+\\ & {\left[1+n(n-1) \frac{\pi^2}{6} \gamma_1(q) \frac{1}{(\ln z)^2}\right.} \\
		& \left.+n(n-1)(n-2)(n-3) \frac{7 \pi^4}{360} \gamma_3(q) \frac{1}{(\ln z)^4}+\cdots\right],
	\end{aligned}
\end{equation} 
where
\begin{equation}\label{Expansion}
	\gamma_n(q)=\int_0^{\infty} \mathrm{d} x \frac{x^n}{2 \ln q} \ln \left[\frac{\exp (x)+q}{\exp (x)+q^{-1}}\right] / \int_0^{\infty} \mathrm{d} x \frac{x^n}{\exp (x)+1},
\end{equation}
and it can be proved that $\gamma_n(q)>1$ when $q \ne 1$  and $\gamma_n(q)=1$ for $q =1$.
Substituting the expansion of Eq.\eqref{Expansion} to Eq. \eqref{TF3} leads to,
\begin{equation}\label{TF_model_3}
\begin{aligned}
\nabla^2 \phi(\mathbf{r})& = 4 \pi e \frac{g}{h^3\lambda^3}
\left\{%
\left(
\frac{(\ln \tilde{z})^n}{\Gamma(n)}+ \left[1+n(n-1) \frac{\pi^2}{6} \gamma_1(q) \frac{1}{(\ln \tilde{z})^2}\right.\right. \right.\\
& \left. \left.+n(n-1)(n-2)(n-3) \frac{7 \pi^4}{360} \gamma_3(q) \frac{1}{(\ln \tilde{z})^4}+\cdots\right]\right)-\\
&\left(%
\frac{(\ln z)^n}{\Gamma(n)}+ \left[%
	1+n(n-1) \frac{\pi^2}{6} \gamma_1(q) \frac{1}{(\ln z)^2}\right. \right.\\
& \left. \left. \left. +n(n-1)(n-2)(n-3) \frac{7 \pi^4}{360} \gamma_3(q) \frac{1}{(\ln z)^4}+\cdots\right]\right)\right\}\\
&-4 \pi Q \delta(\bf{r}) 
\end{aligned}%
\end{equation}
where $ln\tilde{z}=\beta (\mu-e\phi)$. 

Let's assume that the charge produces only a linear response meaning that, $|e\phi/\mu| \ll 1$, One can therefore expand different powers of $ln \tilde{z}$ up to linear terms of $\phi$ Eq. \eqref{TF_model_3} leads to,
\begin{equation}
	\nabla^2\phi(\mathbf{r})=\frac{3}{2}\left[\lambda_{F}^{(q)}\right]^{-2} \phi-4\pi Q\delta(\mathbf{r}),
\end{equation}
whose solution is
\begin{equation}
\phi=\frac{Q}{r}exp{\left(-\sqrt{\frac{3}{2}}\frac{r}{\lambda_{F}^{(q)}}\right)}
\end{equation}
where
\begin{equation}
	\left[\lambda_{F}^{(q)}\right]^{-2}=\frac{4\pi n_0 e^2(\beta \mu)^{3/2}(1-\frac{1}{3} \vartheta_1^q(\beta \mu)-\frac{5}{3} \vartheta_2^q(\beta \mu))}{\mu}.
\end{equation}
Here we have 
\begin{equation}
	\begin{array}{l}
		{n_0} = \frac{1}{{3{\pi ^2}}}{\left( {\frac{{2m\mu }}{{{\hbar ^2}}}} \right)^{\frac{3}{2}}}\\
		\\
		\vartheta _1^q(\beta \mu ) = \frac{{{\pi ^2}}}{8}{\gamma _1}(q){\beta ^{ - 2}}{\mu ^{ - 2}}\\
		\\
		\vartheta _2^q(\beta \mu ) = \frac{{7{\pi ^4}}}{{960}}{\gamma _3}(q){\beta ^{ - 4}}{\mu ^{ - 4}}
	\end{array}
\end{equation}
In order to investigate the $q$-deformation effects on the Thomas-Fermi screening length, the quantity of $\alpha$ is defined as,
\begin{equation}
	\alpha=\frac{\lambda_{F}^{q}}{\lambda_{F}}=\sqrt{\frac{1-\frac{1}{3}\frac{\pi^2}{8}(\beta\mu)^{-2}}{1-\frac{1}{3} \vartheta_1^q(\beta \mu)-\frac{5}{3} \vartheta_2^q(\beta \mu)}}
\end{equation}
where $\lambda_{F}$ is the screening length for the ordinary fermions \cite{Shivamoggy}.

\begin{figure}[hbt!]
	\includegraphics{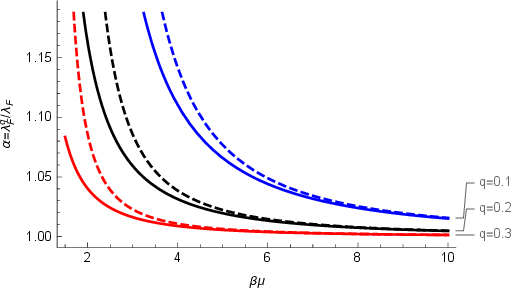}
	\caption{The ratio of $\alpha=\lambda_{F}^{q}/\lambda_{F}$ for $q=0.1, 0.2$ and $0.3$. The dashed curve shows $\alpha$ up to $\gamma_1(q)$ corrections and the solid curve shows the value up to $\mathcal{O}(\beta \mu)^{-4}$ because of $\gamma_3(q)$ correction. }\label{fig1}
\end{figure}

\begin{figure}[hbt!]
	\includegraphics{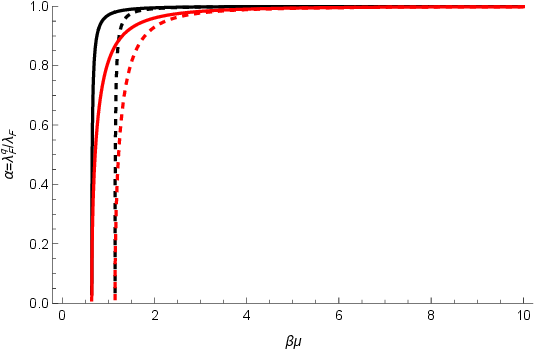}%
	\caption{Same as \figurename{\ref{fig1}} but $q=0.4$ (Black) and $q=0.6$ (Red).}\label{fig2}
\end{figure}

In \figurename{\ref{fig1}}, $\alpha$ has been plotted for values of $q=0.2, 0.3\, \text{and}\, 0.4$. In this figure the solid curves consider the $\mathcal{O}(\frac{\mu}{k_BT})^{-2}$ involved $\gamma_1(q)$ term and dashed curves show the  $\mathcal{O}(\frac{\mu}{k_BT})^{-4}$ related to $\gamma_3(q)$ correction. \figurename{\ref{fig2}} is the same as \figurename{\ref{fig1}} but for higher values of $q$. As it is clear from these figures, the $q$-deformation has more effects on the screening length  at higher temperatures and decreasing temperature, decreases the effect of $q$ statistics. On the other hand, one can see that considering $\gamma_3(q)$ corrections is considerable at lower temperatures and decrease the effects of $q$-deformation. 

As an interesting result, our calculations also show that, at higher temperatures, the behavior of $q$-screening length in Thomas-Fermi model is different for various values of $q$, namely $q$-deformation leads to increase in screening length ($\alpha>1$) at $q\lesssim 0.3$. The results also show that $q$-deformed screening length is smaller than ordinary one higher values of $q$ (for $q>0.3$, $\alpha < 1$) and it is clear from figures that the ratio of $q$-screening length and ordinary one tends to $1$ at sufficiently low temperatures. This means that all $q$-corrections, therefore, vanish at zero temperature. It is be mentioned that $q > 1$ values did not considered here because of 
the symmetry property of $q$-deformed distribution function, $f_{k,q}=f_{k,q^{-1}}$. 

%
%
%
\section {Summary and Conclusion}
The Thomas-Fermi screening length has been investigated in $q$-deformed statistics. The calculations were made based on the $q$-deformed generalized Fermi integral at finite temperature. 
We found that the screening length of $q$-fermions is greater than ordinary fermions for some values of deformation parameter. Some other values of $q$, on the other hand, decrease the
screening length. The results showed that deformation effects vanish at zero temperature limit. It found that 
considering higher correction terms of $q$-deformation, along with temperature dependence, leads to more effects on $q$-screening.
\newpage
\acknowledgements{Financial support from Ilam University research council is gratefully acknowledged.}

\end{document}